\documentclass[%
 aip,
jap,
reprint, %onecolumn %%%%% Comment onecolumn to override one column format %%%%%%%
floatfix,
preprintnumbers
]{revtex4-2}
\usepackage{graphicx}
\usepackage{mathptmx}
\usepackage{etoolbox}

\makeatletter
\def\@email#1#2{%
 \endgroup
 \patchcmd{\titleblock@produce}
  {\frontmatter@RRAPformat}
  {\frontmatter@RRAPformat{\produce@RRAP{*#1\href{mailto:#2}{#2}}}\frontmatter@RRAPformat}
  {}{}
}%

\def\maketitle{
\@author@finish
\title@column\titleblock@produce
\suppressfloats[t]}
\makeatother
\usepackage[english]{babel}

\newlength{\figsize}
\newlength{\subfigsize}
\begin{document}
\setcounter{page}{996}
\title{On the optical field confinement at the anapole mode in nanohole silicon metasurfaces}

\author{A. V. Panov \relax{(https://orcid.org/0000-0002-9624-4303)}}
\affiliation{
Institute of Automation and Control Processes,
Far Eastern Branch of Russian Academy of Sciences,
5 Radio St., Vladivostok, 690041, Russia}
\email{panov@iacp.dvo.ru}
\preprint{Bulletin of the Russian Academy of Sciences: Physics \textbf{88} (6), pp. 996-999 (2024) \doi{10.1134/S1062873824706974}}

\date[Submitted: ]{15 December 2023}%
\accepted[Accepted: ]{29 January 2024}%
\published[Published Online: ]{26 February 2024}

\begin{abstract}
Recently, researchers demonstrated numerous types of the metasurfaces with non-radiating states including the anapole ones.
These metasurfaces have promising properties for various applications in optics which are mainly connected with the optical field confinement. 
But by itself, the existence of the anapole state does not make one available to exploit this feature. 
For example, Ospanova et al. [\textit{Laser \& Photonics Rev.}, 2018, vol. 12, p. 1800005] presented a grouped nanohole array in silicon slab possessing anapole mode in visible range. 
In this work, it is proved that despite of the existence of the anapole mode in the proposed nanostructure the valuable electric field confinement is not reached. 
Thus most expected uses of the anapole state cannot be gained through this nanostructure.
As demonstrated here, the aforementioned nanostructure may be replaced by the square nanohole array having the anapole state with different meta-atom and much larger electric field confinement.
\end{abstract}

\maketitle

% \paragraph*{Keywords:} 

% Author: Please give full first and last names for authors and include * after the name of all corresponding authors

%\author{Andrey V. Panov*}

% Dedication

% \dedication{Optional dedication here. If no dedication is required, please leave blank}

% Affiliations: Please provide adacemic titles (Prof. or Dr.) for all authors where applicable, and include an institutional email address for all corresponding authors
% \begin{affiliations}
%\address{
%Institute of Automation and Control Processes,
%Far East Branch of Russian Academy of Sciences,
%5, Radio st., Vladivostok, 690041, Russia\\
%Email Address: panov@iacp.dvo.ru
%}
% \end{affiliations}

% Keywords: Please provide a minimum of three and a maximum of seven keywords, separated by commas

% \keywords{Anapole state, Dielectric metasurface, Freestanding metasurface}

% \maketitle

Nowadays, optical metasurfaces are of great interest as a promising platform for light manipulation \cite{Su18,Kamali18,Valitova22,Kharitonov22}.
A significant portion of studies in the field of the optical metasurfaces focused on nonradiating states.
Research on the nonradiating anapole state in nanophotonics is rapidly developing in recent years.
Most frequently, the anapole state arises from the destructive interference between Cartesian electric dipole and toroidal electric dipole moments of the expansion of induced electric current density inside a subwavelength structure.
At this state, the far-field radiation from the structure is severely suppressed, while inside the latter, the electric field is not equal to zero and even often is largely enhanced (confined).
Typical applications of the anapole states are linked with the field confinement in the nanostructures: the enlarged optical nonlinear effects \cite{Saadabad22}, sensing \cite{Zeng22}, ultranarrow perfect absorption \cite{He22}.
Also the anapole states are employed for switchable electromagnetically induced transparency \cite{Sun20,Stenishchev21}, ultracompact energy transfer in metachains \cite{Huang21}, cloaking \cite{Saadabad22}, and metabolometers \cite{Cojocari23}.
Generally, the maximum enhancement of the nonlinear effects due to the anapole states is two or more orders of magnitude  compared to unstructured materials \cite{Grinblat16,Rocco18,Timofeeva18,Baranov18}.
Such enlargement of the optical nonlinearity requires even higher local amplification of the electric energy as a consequence of its inhomogeneity within the nanostructure.

In this paper, on the example of the grouped nanohole array metasurface proposed in Ref.~\cite{Ospanova18a} it is demonstrated that such a nanostructure is suboptimal for the field confinement regardless of the existence of the anapole mode.
As a measure of the electric energy concentration, the modeling of the nonlinear optical effects (for example, the optical Kerr effect) can be used.
As a result of the electrical field confinement, the effective optical nonlinearity of the nanostructure should be enhanced.
The estimates of the the effective second order refractive index are done with the use of three-dimensional finite-difference time-domain (FDTD) simulations of light propagation through the nanostructure with optical nonlinearity.
The details of the numerical procedure are described in Ref.~\cite{Panov18}.
The size of the FDTD computational domain for simulations is $1.8\times 1.8\times 9.5$~$\mu$m with the space resolution of 2~nm.
The wavelength for the computations of the effective second order refractive index $\lambda=529$~nm was taken from the anapole frequency obtained for the anapole state in Ref.~\cite{Ospanova18a}, the refractive index of silicon at this frequency $n = 4.16$ \cite{Aspnes83} with neglected extinction coefficient.

\begin{figure}[tbh!]
{\centering\includegraphics[width=\figsize]{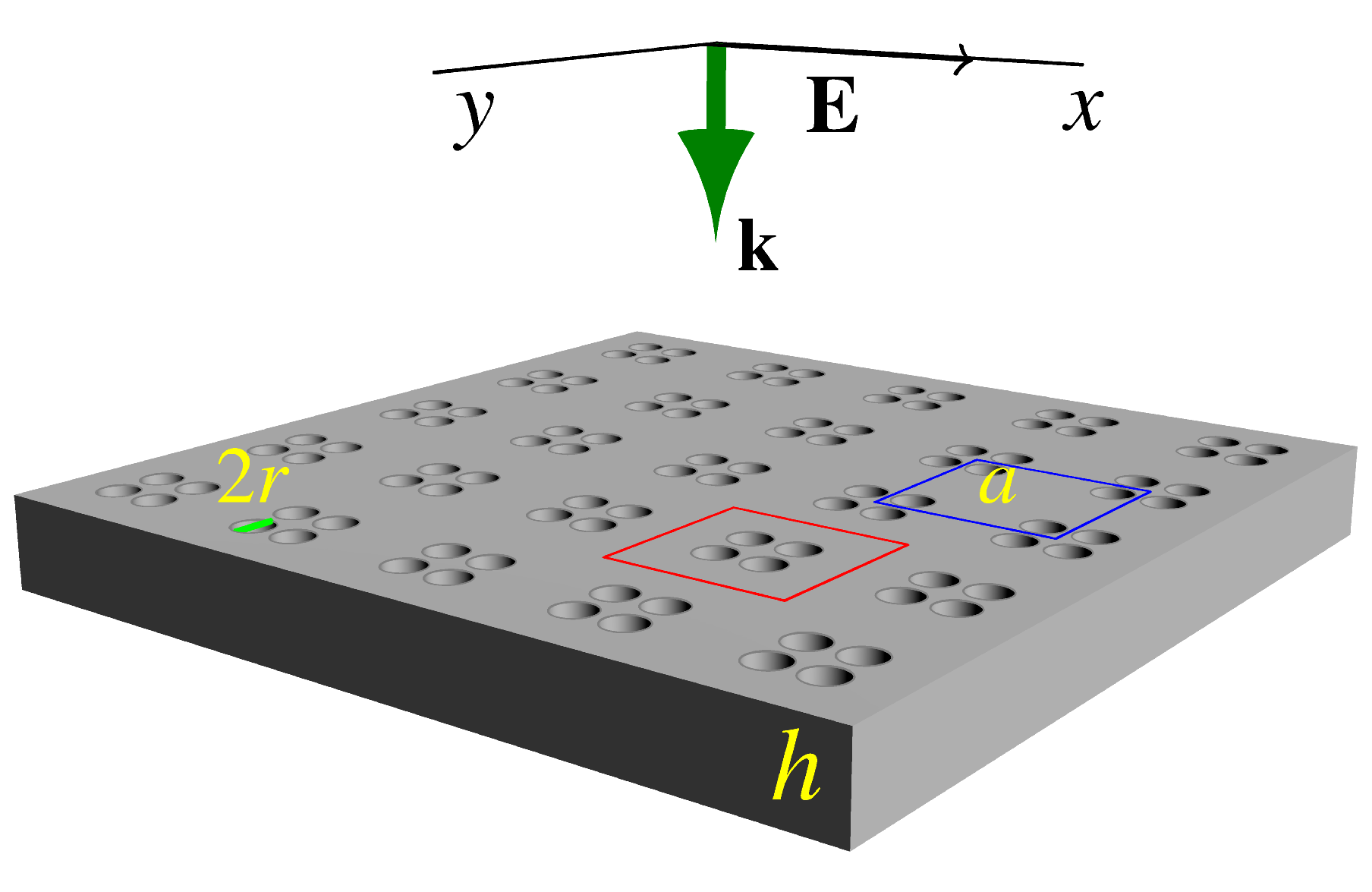}\par} 
\caption{\label{disk_gr_pores}  
Schematic of the simulated freestanding metasurface comprising a lattice array of nanoholes in a silicon slab. 
% The Gaussian beam is incident normally on the metasurface. 
}
\end{figure} 

\begin{figure}[tbh!]
{\centering
\includegraphics[width=\subfigsize]{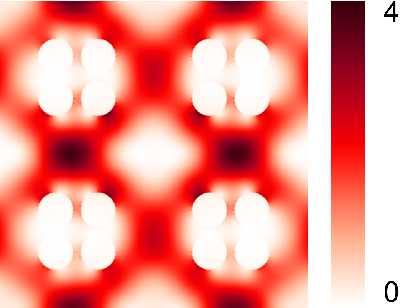}\hspace{1em}
\includegraphics[width=\subfigsize]{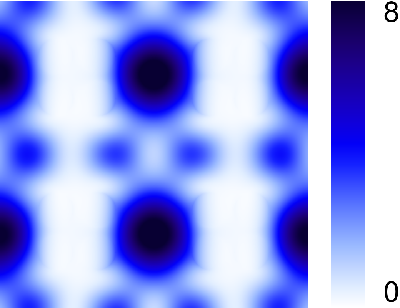}\par}
\caption{\label{ener_dist_Si_400_lattcylhol} 
Time-averaged distributions of electric $|E|^2$ (left part, red color) and magnetic $|H|^2$ (right part, blue color) energy densities in the Si nanohole array with $a=200$~nm, $h=100$~nm, $r=22.5$~nm at $\lambda=529$~nm. 
The energy densities are compared with ones for the unperforated Si slab.
The distributions are calculated within plane $h/2$.
%The incident Gaussian beam is polarized along the vertical direction.
}
\end{figure}

Fig.~\ref{disk_gr_pores} depicts schematics of the the nanohole array metasurface proposed in Ref.~\cite{Ospanova18a} with the red-lined square representing the ``metamolecule'' used there. 
The optimal parameters for the anapole state obtained by Ospanova et al. were as follows: the thickness of the silicon slab $h=100$~nm, the radius of the pore $r=22.5$~nm, the lattice parameter $a=200$~nm, the distance between the centers of the closest pores was 55~nm. 
The time-averaged electric $|E|^2$ and magnetic $|H|^2$ energy distributions at cross-section of the metasurface at $h/2$ are shown for these parameters in Fig.~\ref{ener_dist_Si_400_lattcylhol}. 
The energies in Fig.~\ref{ener_dist_Si_400_lattcylhol} are compared with ones observed in the unperforated Si slab.
As evident, the maximum values of the energies are only several times larger than those for the unpatterned slice. 

The next figure shows results of the modeling of the optical Kerr effect in this nanostructure with the same parameters and varying radius of the nanohole.
The effective second order refractive index of the Si grouped nanohole arrays $n_{2{\mathrm{eff}}}$ is divided by the second order refractive index of the unpatterned slab $n_{2{\mathrm{bulk}}}$.
In the range of the radii $r=20{-}27$~nm, the ratio $n_{2{\mathrm{eff}}}/n_{2{\mathrm{bulk}}}$ steadily varies from 0.8 to 3.
Thus this nanostructure does not exhibit the valuable enhancement of the nonlinear effect.
In other words, a strong field localization is not reached for the structure proposed in Ref.~\cite{Ospanova18a}.
Hence it could hardly be utilized for typical uses of the optical anapole state which require the field confinement while
other applications of the anapole mode are still mostly speculative. 

\begin{figure}[tbh!]
{\centering
\includegraphics[width=\figsize]{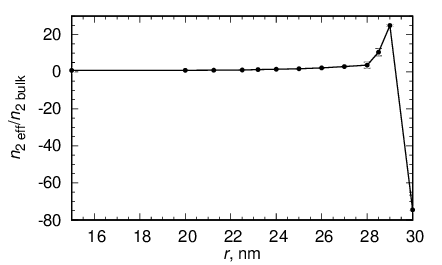}
\par} 
\caption{\label{n2_r_Si_400_lattcylhol_a2_h1_n2}
Relative enhancement of the effective second order refractive index of the Si grouped nanohole arrays with $h=100$~nm, $a=200$~nm at $\lambda=529$~nm as a function of the nanohole radius $r$.
}
\end{figure} 

\begin{figure}[tbh!]
\begin{center}
{\centering
\includegraphics[width=\figsize]{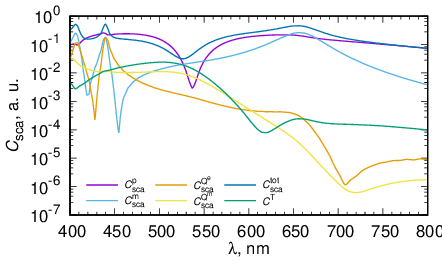}
\par
} 
\end{center}
\caption{\label{Csca_Si_approx_toroidalz_cylcen_log}
Scattering cross section spectra for the multipole contributions (electric dipole $C_{\mathrm{sca}}^{\mathrm{p}}$, magnetic dipole $C_{\mathrm{sca}}^{\mathrm{m}}$, electric quadrupole $C_{\mathrm{sca}}^{\mathrm{Q^e}}$, magnetic quadrupole $C_{\mathrm{sca}}^{\mathrm{Q^m}}$ ), their sum $C_{\mathrm{sca}}^{\mathrm{tot}}$ and the electric toroidal dipole moment $C^{\mathrm{T}}$ for a lattice element of the array of the overlapping circular nanoholes in Si with $a=200$~nm, $h=100$~nm, $r=30$~nm. 
Refractive index is assumed to be constant over the whole wavelength range.
}
\end{figure} 

\begin{figure}[tbh!]
\begin{center}
{\centering
\includegraphics[width=\figsize]{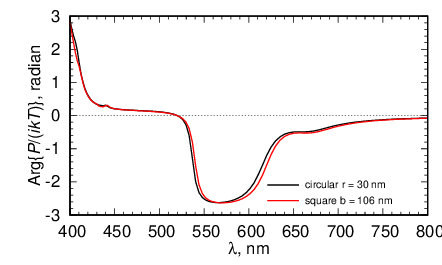}
\par
} 
\end{center}
\caption{\label{Phi_del_cirsqr}
Difference of the phases between electric dipole mode $P$ and toroidal dipole mode $T$ for the lattice elements of arrays of overlapping circular and square nanoholes in Si slab with $a=200$~nm, $h=100$~nm: grouped circular with $r=30$~nm and square with side $b=106$~nm calculated. 
Refractive index is assumed to be constant over the whole wavelength range.
}
\end{figure}

\begin{figure}[tbh!]
{\centering\includegraphics[width=\figsize]{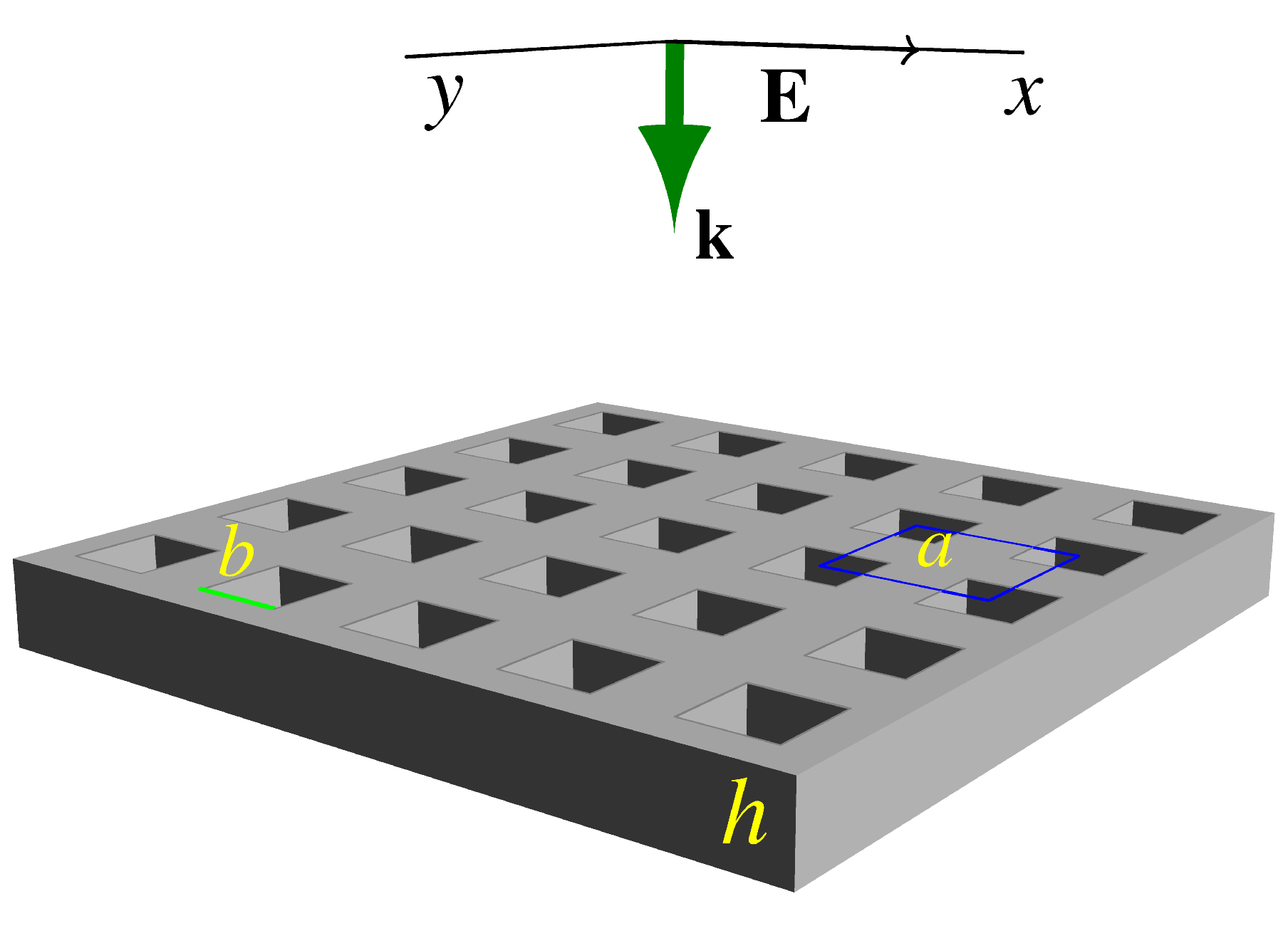}\par} 
\caption{\label{square_pore5}  
Schematic of the freestanding metasurface comprising a square lattice array of the square nanoholes in a silicon slab. 
% The Gaussian beam is incident normally on the metasurface. 
}
\end{figure}

For $r>27.5$~nm, the nanostructure presented in Ref.~\cite{Ospanova18a} becomes invalid as the closest nanoholes overlap.
Interestingly,  $|n_{2{\mathrm{eff}}}|/n_{2{\mathrm{bulk}}}$ rapidly rises up to 70 for the nanostructure comprising overlapping nanoholes.
Fig.~\ref{Csca_Si_approx_toroidalz_cylcen_log} illustrates a multipole decomposition of scattering cross sections up to second order multipoles for a lattice element delineated by the blue-lined square in Fig.~\ref{disk_gr_pores}.
This decomposition is performed using approximate equations from Ref.~\cite{Alaee18} which coincide with ones utilized in Ref.~\cite{Ospanova18a}.
The total scattering cross section $C_{\mathrm{sca}}^{\mathrm{tot}}$ and the electric dipole cross section $C_{\mathrm{sca}}^{\mathrm{p}}$ display minima near $\lambda=529$~nm, while the toroidal electric dipole cross section $C^{\mathrm{T}}$ reaches a maximum in the close range.
The anapole state is observed when the phases of the electric dipole  moment $\mathbf{P}$ and $-ik\mathbf{T}$ coincide, where $\mathbf{T}$ is toroidal electric dipole moment and $k=2\pi/\lambda$.
This phase difference between $\mathbf{P}$ and $-ik\mathbf{T}$ is represented in Fig.~\ref{Phi_del_cirsqr}.
As obvious, the anapole condition for a lattice element delineated by the blue lines in Fig.~\ref{disk_gr_pores} when the curve for the overlapping circular nanoholes in Si plate crosses zero is seen near the wavelength of interest.
Therefore this metasurface has different meta-atom.
The array of overlapping circular nanopores can be replaced with the simpler metasurface comprising the square lattice of the square nanoholes which is illustrated by Fig.~\ref{square_pore5}.

The results of the multipole decomposition of scattering cross sections for the lattice element of the square nanohole array with the side size $b=106$~nm delineated by the blue lines in Fig.~\ref{square_pore5} almost coincide with ones represented in Fig.~\ref{Csca_Si_approx_toroidalz_cylcen_log}.
The phase difference between the electric dipole and toroidal electric dipole moments for the square nanopores is displayed in  Fig.~\ref{Phi_del_cirsqr}.
The enhancement of the optical Kerr effect for the square nanohole array in Si slab with  $a=200$~nm, $h=100$~nm and  $b=106$~nm at $\lambda=529$~nm $n_{2{\mathrm{eff}}}/n_{2{\mathrm{bulk}}}\approx 76$.
The effective second order refractive $n_{2{\mathrm{eff}}}$ index rapidly arises in proximity to the anapole state and may change the sign~\cite{Panov22}.
More information on the anapole state and the optical nonlinearity of the square nanohole arrays can be found in Ref.~\cite{Panov23,Panov22a}.
Consequently, this nanostructure possessing the anapole state show much larger electric energy concentration as compared with one presented in Ref.~\cite{Ospanova18a}.

In summary, the existence of the anapole state does not guarantee the electric energy confinement within the nanostructure.
The modeling of the optical nonlinearity presented in this study shows that the  grouped nanohole array metasurface displays very low enhancement of the effective optical Kerr effect and as a consequence a low energy concentration.
In simulations, the energy enhancement should be checked by calculating the energy distributions inside the nanostructure and comparing them with unpatterned material. 
Nevertheless, this metasurface can be replaced by the square nanohole array in the silicon slab which displays much higher optical nonlinearity as a result of the field confinement.
% The conclusions of Ref.~\cite{Ospanova18a} about strong field localization in their structure and possibility to use it for applications requiring this localization are not justified.
Therefore, more search for optimal nanostructures in the similar works should be performed.

%\medskip
%\textbf{Declaration of Competing Interest} \par\nopagebreak 
%The author declares that he has no known competing financial interests or personal relationships that could have appeared to influence the work reported in this paper.

\section*{Conflict of interest}
The author declares that he has no conflicts of interest.

%\medskip
%\textbf{Data availability} \par\nopagebreak 
%Data will be made available on request.

\section*{Acknowledgements}
The results were obtained with the use of IACP FEB RAS Shared Resource Center ``Far Eastern Computing Resource'' equipment (https://www.cc.dvo.ru).

\bibliography{nlphase}

\end{document}